\documentstyle [12pt]{article}
\begin{document}
\baselineskip .3in
\begin{titlepage}
\begin{center}{\large {\bf Statistical Physics of the Travelling Salesman 
Problem}}$~^{*}$
\vskip .2in
{\bf Anirban Chakraborti}$~^{(1)}$ and
{\underline {\bf Bikas K. Chakrabarti}}$~^{(2)}$\\
{\it Saha Institute of Nuclear Physics},\\
{\it 1/AF Bidhan Nagar, Calcutta 700 064, India.}\\
\end{center}
\vskip .3in
{\centerline{\bf Abstract}}
If one places $N$ cities randomly on a continuum in an unit area, extensive 
numerical results and their analysis (scaling, etc.) suggest that the best 
optimized travel distance per city becomes $l_E \simeq 0.72/\sqrt N$ for the 
Euclidean metric, and $l_M \simeq 0.92/\sqrt N$ for the Manhattan metric. The 
analytic bounds, we discuss here, give 
$0.5< l_E\sqrt N < 0.92$ and $0.64 < l_M\sqrt N < 1.17$. When 
the cities are randomly placed on a lattice with concentration $p$,
we find (with $N=p$ for unit area of the country)
$l_E\sqrt p$ and $l_M\sqrt p$ vary monotonically with $p$: $l_E\sqrt p =
l_M\sqrt p = 1$ for $p=1$, and $l_E\sqrt p\simeq 0.72$ and $l_M\sqrt p \simeq 
0.92$ as $p\rightarrow 0$. The problem 
is trivial for $p=1$ but it reduces to the continuum TSP for $p\rightarrow 0$. 
We did not get any irregular behaviour at any intermediate point, e.g., the 
percolation point.
The crossover from the triviality to the NP- hard problem seems to occur at 
$p<1$.
\vskip 3in
\noindent
---------------------------------------------------------------------------\\
\noindent
* K. C. Kar Memorial Lecture, 1999 (to be published in Indian J. of Theo.
Phys., Calcutta).
\end{titlepage}
\newpage
\noindent
{\large {\bf 1. Introduction}}
\vskip .1in
\noindent
In everyday life we face several complex problems, classified as
combinatorial optimization problems, the solutions of which are of great practical importance.
Research in this area tries to find different efficient techniques for finding 
the extremum (maximum or minimum) values of a function of many different 
independent variables [1-3].

The travelling salesman problem (TSP) is a simple example of a combinatorial 
optmization problem and perhaps the most famous one. Given a certain set of 
cities and the intercity distance metric, a travelling salesman must find the 
shortest tour in which he visits all the cities and comes back to his starting 
point. It is a non-deterministic polynomial complete (NP- complete) problem.
NP problems are those for which a potential solution can be checked efficiently
for correctness, but finding such a solution appears to take time which scales 
exponentially with the size $N$ in the worst case. The completeness property of
 NP- complete problems means that if it is possible to find a deterministic 
algorithm that solves {\it one} NP- complete problem in polynomial time, then 
the other NP- complete problems could also be solved in polynomial time.

In the TSP, the most naive algorithm for finding the optimal tour
would have to consider all the $(N-1)!/2$ possible tours for $N$ number of 
cities and check for the shortest of them.
Working this way, the fastest computer available today would require more time 
than the current age of the universe to solve a case with about $30$ cities. 
The typical-case behaviour is difficult to characterize for the  
the TSP though it is believed to require exponential time to 
solve in the worst case. For this reason
the TSP serves as a prototype problem for the study of the 
combinatorial optimization problems in general.

In the normal TSP, we have $N$ number of cities distributed in some continuum 
space and we determine the average optimal
 travel distance per city $\bar l_E$ in the Euclidean metric (with 
$\Delta r_E= \sqrt {\Delta x^2+\Delta y^2}$),
or $\bar l_M$ in the Manhattan metric (with $\Delta r_M= |\Delta x|+|\Delta y|$).
Since the average distance per city (for fixed area) scales with the number of cities $N$ 
as $1/ \sqrt N$, we find that the normalized travel distance per 
city $\Omega_E=\bar l_E \sqrt N$ or
$\Omega_M=\bar l_M \sqrt N$ become the optimized constants and their values
depend on the method used to optimize the travel distance. In section 2, 
we discuss
some algorithms used to determine the optimal tour and find the values of the
constants $\Omega_E$ and $\Omega_M$ for the optimized travel. In section 3, we 
present an analytic method to estimate the upper and lower bounds of $\Omega_E$ and $\Omega_M$.

In the lattice version of the TSP, the cities are represented by randomly 
occupied lattice sites of a two- dimensional square lattice; the fractional 
number of occupied sites being $p$ (lattice occupation concentration). 
In this case the average optimal travel distance in the Euclidean metric 
$\bar l_E$, and in the Manhattan metric $\bar l_M$, vary with the lattice 
concentration $p$. Then the normalised travel distance per city are defined as
$~\Omega_E=\bar l_E ~\sqrt p~$ and $~\Omega_M=\bar l_M ~\sqrt p~$. 
In section 4, we study the variation of $\Omega_E$ and
$\Omega_M$, and the ratio $\Omega_M / \Omega_E$ with $p$. 
Finally, we draw conclusions in section 5.

\vskip .3in
\noindent
{\large {\bf 2. Some heuristic algorithms}}
\vskip .1in
\noindent
The most naive method to obtain an approximate solution of travelling salesman 
problem is the ``greedy" heuristic algorithm [1, 2]. 
Suppose we have a random arrangement of $N$ cities in a square (country) of 
fixed area (taken to be unity). Let us think of any tour to start-with and then
make a local exchange of a pair of cities in the tour. We compute the new tour 
length and if it is lower than the previous one, then the greedy algorithm
accepts the new tour as the starting point for further such modifications. The
``Lin- Kernighan" algorithm [4, 5] considers local exchange between three or 
more cities.

The essential drawback of such local search algorithms is the obvious one of 
getting stuck at a local minimum, where any local rearrangement in the tour does
not improve the optimized tour length.
The ``simulated annealing method" [2, 6] is an ingenious method in analogy with 
the thermodynamic way of avoiding such local minima in free energy (glass 
formations) and achieving the global minimum of a many-body system by {\it slow
cooling} or {\it annealing}. The rapid quenching of the system leads to the 
trapping of the system in a local minimum (or glass) state. The system cannot 
get out of it, since the Boltzmann probability to get out of the minimum drops 
to zero, as the temperature becomes zero due to quenching. This is similar to 
the greedy or other local search algorithms. In the annealing, the system is 
slowly cooled so that as the system falls in a local trap, the finite Boltzmann 
probability ($\sim \rm {exp}(E^{\prime} -E)/kT$, for trap energy $E$ and barrier 
height $E^{\prime} $) allows the system to get out of the trap, maintaining a 
general flow to lower energy states as temperature decreases. Eventually the 
system anneals to the ground state at the lowest temperature.

In the TSP case, one takes the total tour length $L~~(=Nl)$ as the energy $E$ and one 
introduces a fictitious temperature $T$. Initially one takes $T$ very high such that the average total tour length $\bar L$ is much higher than the global 
minimum. The tours are then modified locally and the modified tours are accepted
 with probability $\sim \rm {exp}(\Delta
L/kT)$ where $\Delta L$ is the change in the tour length. In greedy algorithm 
the probability is unity for negative $\Delta L$ and it is zero for positive
$\Delta L$ cases. Here, probability is non-vanishing  
even for $\Delta L$ positive as long as the temperature is nonzero!

Simulated annealing and numerous heuristic generalizations of the local search
algorithm optimize very effectively on small scales involving a small number 
of variables, but fail for the larger scales that require the modification of 
many variables simultaneously.  To deal with the large scales, 
``genetic algorithms" [7] use a ``crossing" procedure which takes two good 
configurations $-$ ``parents", from a population and finds sub-paths that are 
common to the parents. It generates a ``child" by reconnecting
those sub-paths, either randomly or by using large parts of its parents. A 
population of configurations is evolved from one generation to the next using 
these crossings followed by a selection of the best children. However, this 
approach supposedly does not work well in practice since it is extremely 
difficult to produce two parents and cross them to make a child as good as 
them. This is a major drawback of the genetic
algorithms and is responsible for their limited use.

So far, careful analysis of the numerical results obtained indicates that
$\Omega_E \simeq 0.72$ [8] for TSP on continuum.

\vskip .3in
\noindent
{\large {\bf 3. Some analytical results for the bounds for $\Omega$}}
\vskip .1in
\noindent
Although the TSP problem is a multivariable optimization problem (real number of
 variables $\sim N!$ in an $N$ city problem), we now look for an approximate 
analytical solution (upper bound) by expressing the travel distance as a 
function of a single variable and optimizing the distance with respect to 
that variable [9]. As is obvious, the problem is trivial in one dimensional
 case where any directed tour will solve it. In two dimensions, one can again reduce it
(approximately) to an one dimensional problem, where the square (country) is 
divided into strips of width $W$ and within each strip, the salesman visits the
 cities in a directed way. The total travel distance is then optimized with 
respect to $W$.

Let the strip width be $W$ and the probability density of cities be $p~~(=N$ for
unit area).  We have a city at $(0,y_1)$ [See Fig. 1].
The probability that the next city is between distances $x$ and $x+\Delta x$, 
is $p W\Delta x$. The probability that there is no city in the distance 
$x=n\Delta x$, is $(1- p W\Delta x)^n \sim e^{- (p Wx)}$. The probability 
that there is a city between $y$ and $y+\Delta y$, is $\Delta y/W$. Hence the 
probability that there is no other city within distance $y$ is $(1-y/W)$. 
The average distance between any two consecutive cities is therefore 
$$\bar l_E=2\int_{x=0}^{\infty} \int_{y=0}^W \sqrt {x^2+y^2} ~~p Wdx~e^{-(p Wx)} 
\frac {dy}{W}(1-\frac {y}{W})~.\eqno (1) $$ 
The factor $2$ arises to take care of the fact that $y$ can be both positive and 
negative. We make the substitutions: 
$u=p Wx$ and $v=y/W$, so that
$$\bar l_E=2\int_{u=0}^{\infty} \int_{v=0}^1 \frac{1}{p W}
\sqrt {u^2+{p}^2W^4v^2}~~ e^{-u}(1-v)dudv~. $$
We introduce two dimensionless quantities 
$\Omega_E=\sqrt p~~ \bar l_E$ and
$\tilde{W}=\sqrt p ~~W$, so that 
$$\Omega_E=\frac{2}{\tilde{W}}\int_{u=0}^{\infty} \int_{v=0}^1\sqrt {u^2+
{\tilde{W}}^4v^2}~~ e^{-u}(1-v)dudv~.\eqno (2) $$
Using the method of Monte Carlo integration to evaluate the above 
integral, we get the minimum $\Omega_E\sim 0.92$ at normalized strip width
$\tilde {W}\sim 1.73~~$ [See Fig. 2].

In the Manhattan metric the average distance between any two consecutive cities is 
$$\bar l_M=2\int_{x=0}^{\infty} \int_{y=0}^W (x+y) p Wdxe^{-(p Wx)} \frac 
{dy}{W}(1-\frac {y}{W})~. \eqno (3) $$
As before we introduce $u=p Wx$ and $v=y/W$, so that
$$\bar l_M=2\int_{u=0}^{\infty} \int_{v=0}^1 \frac{1}{p W}(u+{p}W^2v) 
e^{-u}(1-v)dudv~, $$
and then introduce the dimensionless quantities
$\Omega_M=\sqrt p ~~\bar l_M$ and
$\tilde{W}=\sqrt p ~~W$, so that 
$$\Omega_M=\frac{2}{\tilde{W}}\int_{u=0}^{\infty} \int_{v=0}^1(u+{\tilde{W}}^2
v) e^{-u}(1-v)dudv~. \eqno (4) $$
Using the method of Monte Carlo integration, we get the
minimum $\Omega_M\sim 1.15$ at the normalized strip width 
$\tilde {W}\sim 1.73~~$ [See Fig. 3].

Note that the relation
$$\Omega_M\simeq \frac {4}{\pi} \Omega_E$$ 
can be explained as follows.
Let $$x=l_E \sin \theta ~~~~~{\rm and} ~~~~~y=l_E \cos \theta~.$$
Then,
$$l_M=x+y=l_E(\cos \theta + \sin \theta)~.$$
Since $\langle x \rangle =\langle y \rangle$,
$$\bar l_M =2 \bar l_E \langle \cos \theta \rangle~.$$
We have now
$$\langle \cos \theta \rangle=\frac{2}{\pi} \int_0^{\pi /2} \cos \theta d\theta 
= \frac{2}{\pi}[\sin \theta {]}_0^{\pi /2}
= \frac{2}{\pi}~.$$
Hence $$\bar l_M =\frac {4}{\pi} \bar l_E ~,~~~~~\rm {or}~~~~~~~~
 \Omega_M=\frac{4}{\pi}\Omega_E~. \eqno (5) $$

Let us now estimate the lower bound of the minimum travel distance per city.
Let the distance between any two cities be denoted by $l$. Then the probability
 that there is a city between $l$ and $l+dl\sim (p-1)2\pi l~dl\sim 2p\pi l~dl$. 
Now, the probability that there is no other city in the distance 
$l~\sim {(1-\pi l^2)}^{p-2}\sim e^{-(p-2)\pi l^2}\sim e^{-p\pi l^2}$.
Therefore, $P(l)dl=(2p\pi l)e^{-p\pi l^2}dl$.
Note that $\int P(l)dl=1$.  Hence the average distance is
$$\bar l_E =\int_0^{\infty}lP(l)dl=2p\pi \int_0^{\infty}l^2e^{-\pi pl^2}dl=
\frac {1}{2}~\frac {1}{\sqrt p}~. \eqno (6) $$
Therefore, the lower bound for $\Omega_E=1/2~$.
The lower bound for $\Omega_M$ can then easily be estimated to be $2/\pi$ in a similar manner.

\vskip .3in
\noindent
{\large {\bf 4. The TSP on randomly diluted lattices}}
\vskip .1in
\noindent

The lattice version of the TSP was first studied by Chakrabarti [10].
In the lattice version of the TSP, the cities are represented by randomly 
occupied lattice sites of a two- dimensional square lattice ($L \times L$), 
the fractional 
number of sites occupied being $p$ (lattice concentration) [11-13]. 
In this case, the average optimal travel distance in the Euclidean metric 
$\bar l_E$, and in the Manhattan metric $\bar l_M$, vary with the lattice 
concentration $p$. We intend to study in this case the variation of the 
normalised travel distance per city, 
$~\Omega_E=\bar l_E \sqrt p$ and $\Omega_M=\bar l_M \sqrt p$, 
with the lattice occupation (city) concentration $p$.

We generate the randomly diluted lattice configuration following the standard
Monte Carlo procedure for 64($=N$) randomly positioned (on the lattice) cities. 
We vary the lattice size from $(8\times 8)$ to $(48\times 48)$ so that
 the lattice concentration $p$ varies from $1.000$ to $0.028$. 
For each such lattice configuration, the exact optimum tour [See Fig. 4]
is obtained with the help of the {\it GNU tsp\_ solve} [14].
We then calculate $l_E$ and $l_M$. At each lattice concentration $p$, we take 
different lattice configurations and then obtain the averages, $\bar l_E$ and 
$\bar l_M$. We then determine $~\Omega_E=\bar l_E ~\sqrt p~$ and
$~\Omega_M=\bar l_M ~\sqrt p~$ and study the variation of $\Omega_E$ and
$\Omega_M$, and of the ratio $\Omega_M /\Omega_E$ with $p$. 
We find that $\Omega_E$ has monotonic variation from $1$ (for $p=1$) to a 
constant $\sim 0.79$ (for $p\rightarrow 0$) and $\Omega_M$ has monotonic 
variation from $1$ (for $p=1$) to the constant $1.01$ (for $p\rightarrow 0$) 
respectively. We believe, with bigger $N$ the value of $\Omega_E$ eventually 
reduces to about $0.72$ as in continuum TSP. Results for higher values of $N$
($\simeq 100$) [15] indeed suggest the same. The ratio $\Omega_M /\Omega_E$ 
changes from $1$ to $1.26~~ (\simeq 4/\pi)$, as $p$ varies from $1$ to $0$ 
[See Fig. 5]. We note that the TSP on randomly diluted lattice is certainly a
trivial problem when $p=1$ (lattice limit) as it reduces to the one-dimensional
 TSP (the connections in the optimal tour are between the nearest neighbours 
along the lattice; Hamiltonian walks).
 However, it is certainly hard at the $p\rightarrow 0$ (continuum) limit.
It is clear that the problem crosses from triviality (for $p=1$) to the NP- hard
problem (for $p\rightarrow 0$) at a certain value of $p$. It seems the 
transition occurs at $p<1$. This requires further investigation.

\newpage
\noindent
{\large {\bf 5. Conclusions}}
\vskip .1in
\noindent
  
If one places $N$ cities randomly on a continuum in an unit area, the best 
numerical results and their analysis (scaling, etc.) suggest that the best 
optimized travel distance per city becomes $l_E \simeq 0.72/\sqrt N$ for the 
Euclidean metric and $l_M \simeq 0.92/\sqrt N$ for the Manhattan metric. The 
analytic bounds we discussed in section 3, gives 
$\Omega_E(=l_E\sqrt N) < 0.92$ and $\Omega_M(=l_M\sqrt N) < 1.17$. When 
the cities are randomly placed on a lattice with concentration $p$, as discussed
 in section 4, we find (with $N=p$ for unit area of the country) that
$\Omega_E(p)$ and $\Omega_M(p)$ are monotonically varying with $p$. The problem 
is trivial for $p=1$ where $\Omega_E(p)=\Omega_M(p)=1$ and it certainly reduces
to the continuum TSP discussed before for $p\rightarrow 0$ 
($\Omega_E \simeq 0.72$ and $\Omega_M \simeq 0.92$; although we observed higher
values, viz., $\Omega_E\simeq 0.79$ and $\Omega_M \simeq 1.01$, since $N$ is 
not suffiently large).
The variations of $\Omega$ with $p$ are found to be monotonic without any 
irregular behaviour at any intermediate point like the percolation point, etc.
The crossover from the triviality to the NP- hard 
problem seems to occur at $p<1$. However, this requires further investigation.  

\vskip 0.3in
\noindent
{\bf Acknowledgement} : We are grateful to R. B. Stinchcombe, A. Percus and
O. C. Martin for very useful comments and suggestions.

\newpage
\noindent
{\bf References}\\
\vskip .1in
\noindent
{\it e-mail addresses} :

\noindent
$^{(1)}$anirban@cmp.saha.ernet.in

\noindent
$^{(2)}$bikas@cmp.saha.ernet.in
\vskip .2 in

\noindent
[1] M. R. Garey and D. S. Johnson, {\it Computers and Intractability: A Guide
to the Theory of NP- Completeness} (1979).\\
\noindent
[2] S. Kirkpatrick, C. D. Gelatt, Jr., and M. P. Vecchi, {\it Science},
{\bf 220}, 671 (1983).\\
\noindent
[3] M. Mezard, G. Parisi and M. A. Virasoro, {\it Spin Glass Theory and Beyond} (1987).\\
\noindent
[4] Y. Usami and M. Kitaoka, {\it Int. J. of Mod. Phys. B}, {\bf 11}, 1519 (1997).\\
\noindent
[5] S. Lin and B. W. Kernighan, {\it Oper. Res.}, {\bf 21}, 498 (1973).\\
\noindent
[6] W. H. Press, S. A. Teukolsky, W. T. Vetterling, B. P. Flannery, {\it Numerical Recipes in C, Second Edition}, 444 (1992).\\
\noindent
[7] D. E. Goldberg, {\it Genetic Algorithms in Search, Optimization and Learning} (1989).\\
\noindent
[8] A. Percus and O. C. Martin, {\it Phys. Rev. Lett.}, {\bf 76}, 1188 (1996).\\ 
\noindent
[9] J. Beardwood, J. H. Halton and J. M. Hammersley, {\it Proc. Camb. Phil. Soc.}, {\bf 55}, 299, (1959); 
R. S. Armour and J. A. Wheeler, {\it Am. J. Phys.}, {\bf 51 (5)}, 405 (1983).\\
\noindent
[10] B. K. Chakrabarti, {\it J. Phys. A: Math. Gen.}, {\bf 19}, 1273 (1986).\\
\noindent
[11] D. Dhar, M. Barma, B. K. Chakrabarti and A. Tarapder, 
{\it J. Phys. A: Math. Gen.}, {\bf 20}, 5289 (1987).\\
\noindent
[12] M. Ghosh, S. S. Manna and B. K. Chakrabarti,
{\it J. Phys. A: Math. Gen.}, {\bf 21}, 1483 (1988).\\
\noindent
[13] P. Sen and B. K. Chakrabarti,
{\it J. Phys. (Paris)}, {\bf 50}, 255, 1581 (1989).\\
\noindent
[14] C. Hurtwitz, {\it GNU tsp\_ solve}, available at : 
http://www.cs.sunysb.edu/\~\\
\noindent
algorith/implement/tsp/implement.shtml\\
\noindent
[15] A. Chakraborti and B. K. Chakrabarti ({\it to be published}).\\

\newpage
\noindent
{\bf Figure captions}\\
\vskip .1 in

\noindent
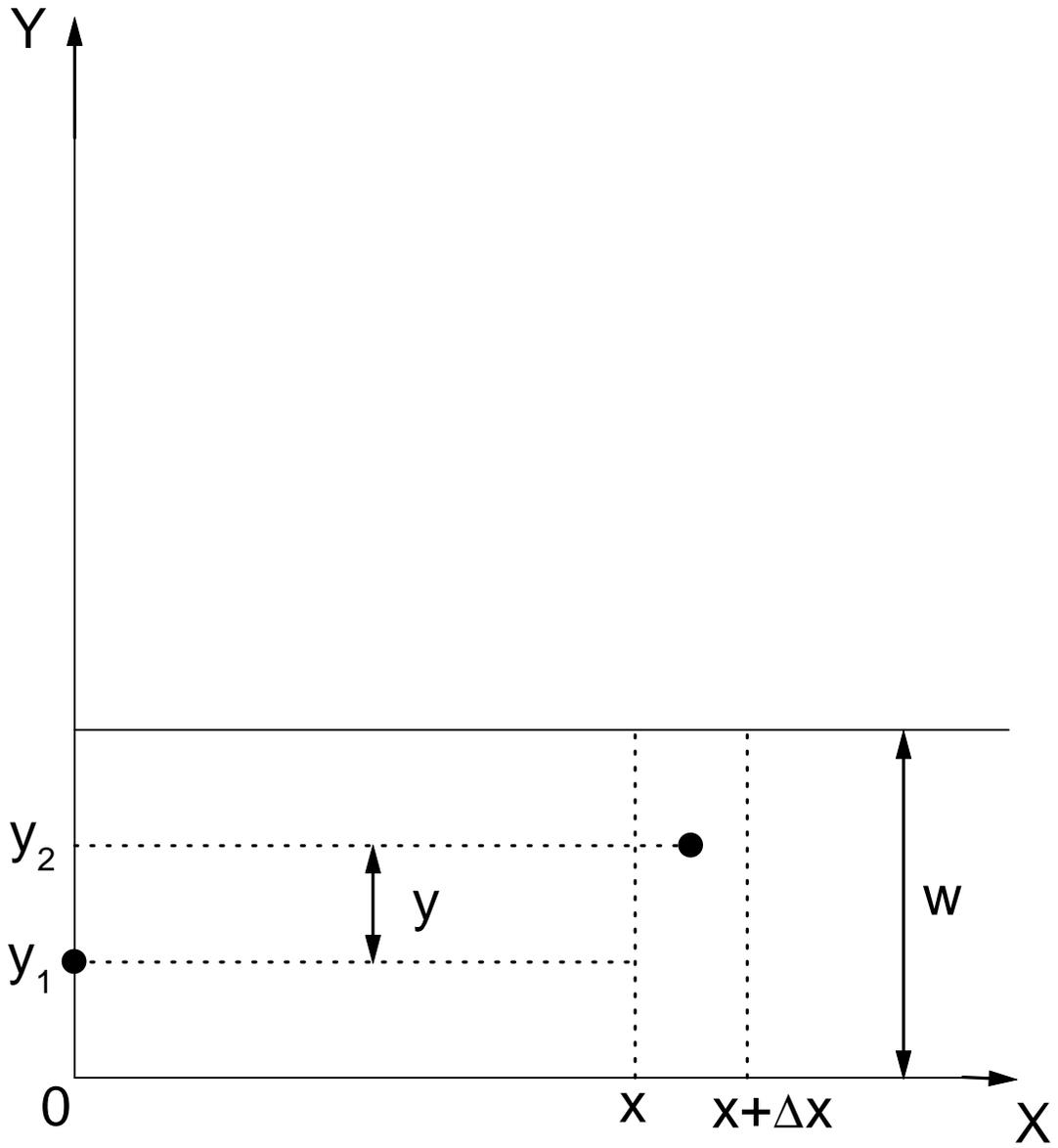
{\bf Fig. 1} : Calculating the average distance between two nearest neighbours 
along a strip of width $W$.\\ 
\noindent
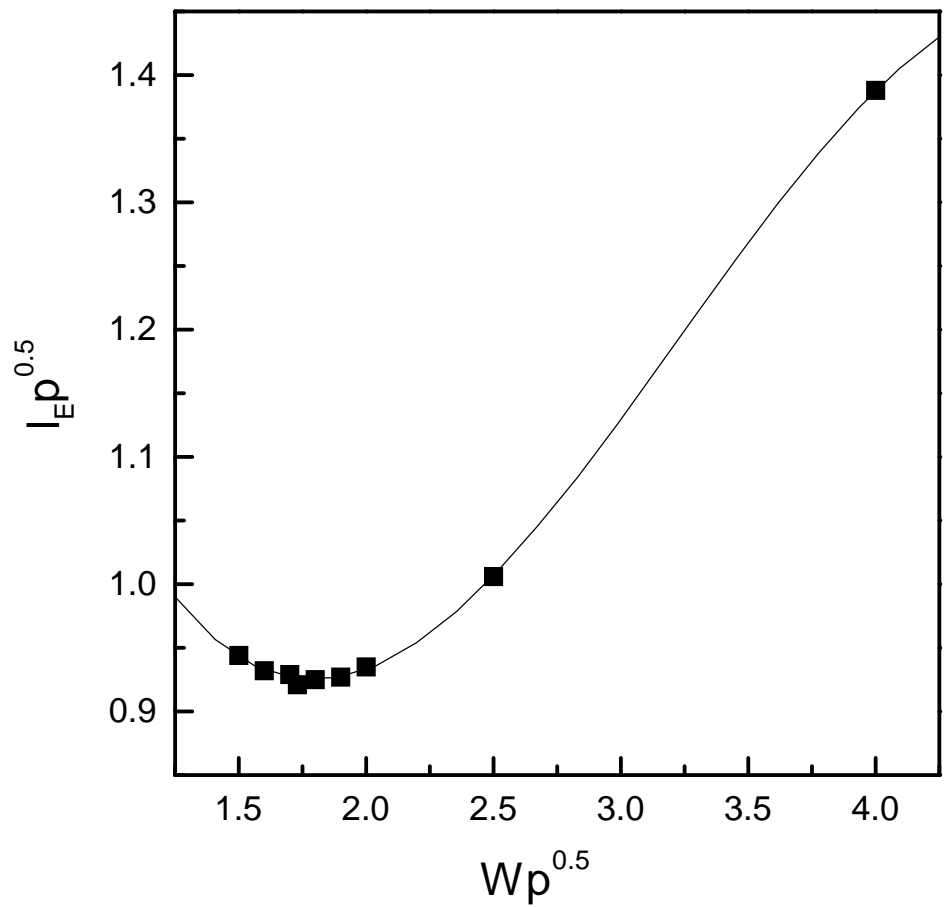
{\bf Fig. 2} : Plot of $l_E \sqrt p$ against $W\sqrt p$ from eqn. (2).\\
\noindent
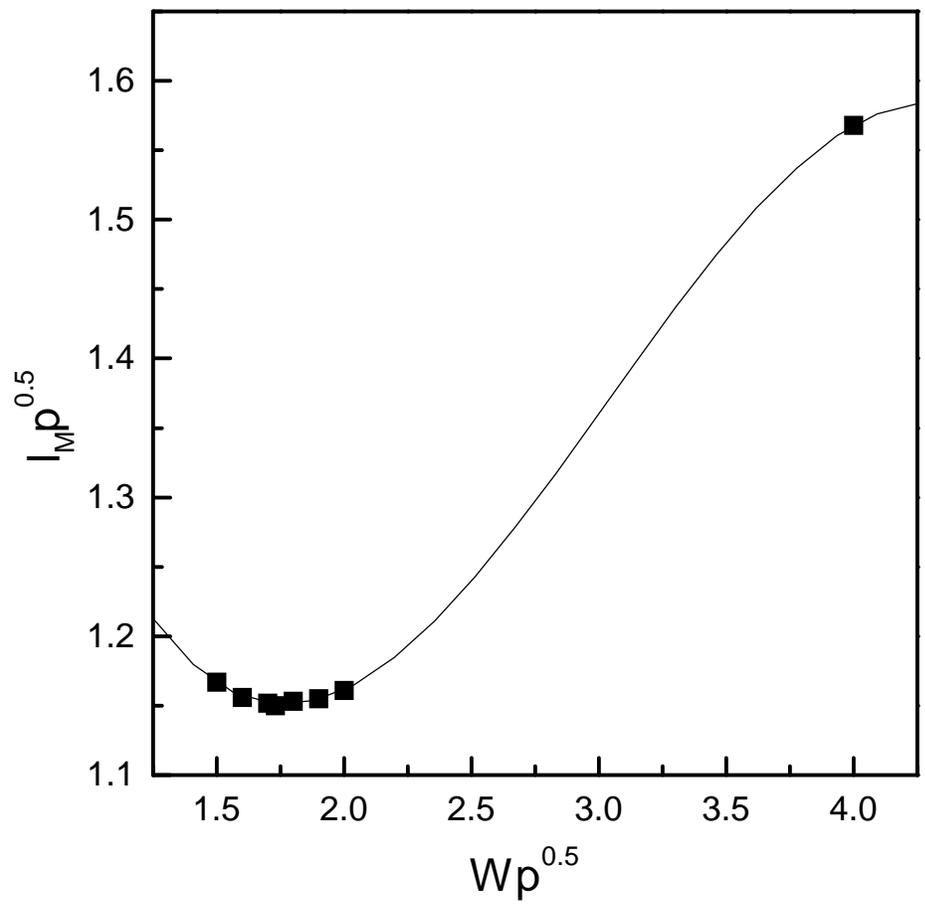
{\bf Fig. 3} : Plot of $l_M \sqrt p$ against $W\sqrt p$ from eqn. (4).\\
\noindent
{\bf Fig. 4} : A typical optimized tour for TSP on dilute lattice in the Euclidean metric for $N=64$ cities.\\
\noindent
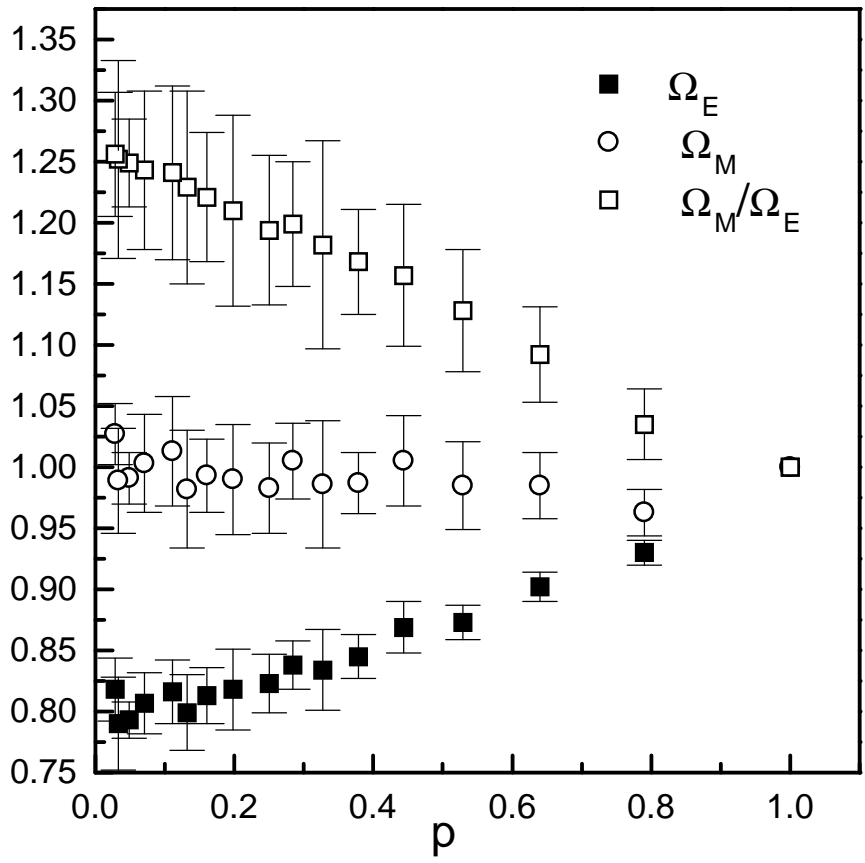
{\bf Fig. 5} : Plot of $\Omega_E$, $\Omega_M$ and $\Omega_M/\Omega_E$ against 
$p$ for TSP on dilute lattice, obtained using the optimization programs (exact)
for $N=64$ cities (fixed).
 The error bars are due to configuration to configuration variations. 

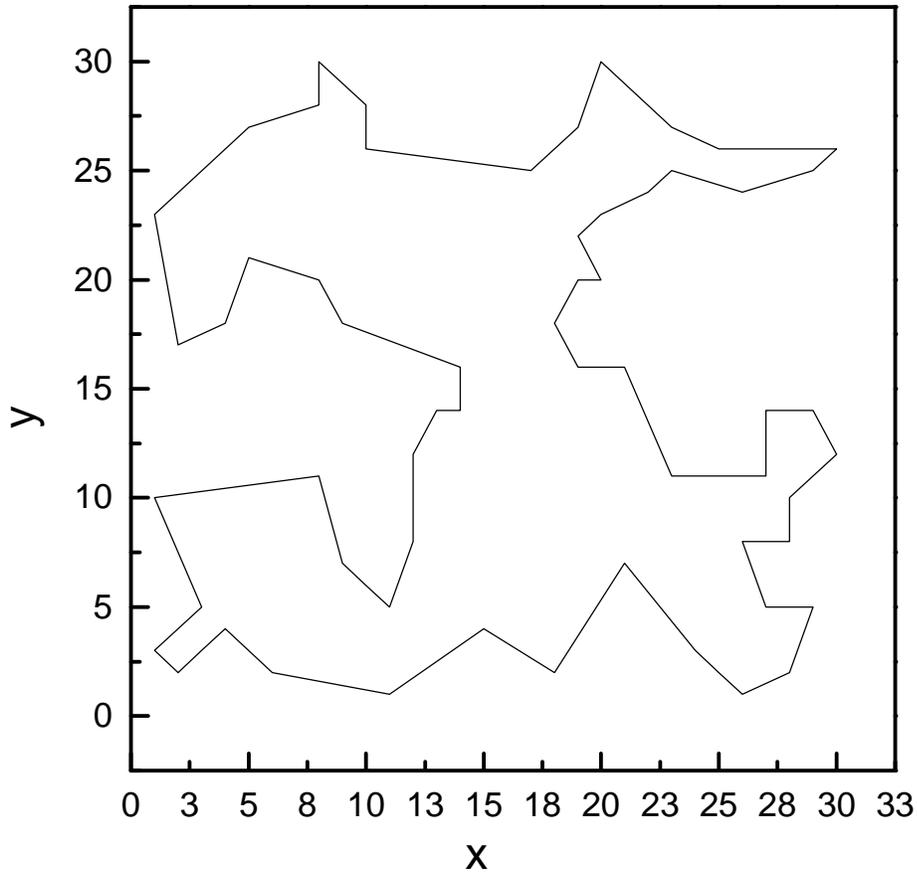
\end{document}